# ASIC IMPLEMENTATION OF DENOISING FILTERS FOR PACEMAKERS

Charan S [1] and Dr. Reena Monica P [2,*]

[1] Student, School of Electronics Engineering, Vellore Institute of Technology, Chennai; charan.s2019@vitstudent.ac.in

[2] Associate Professor and Associate Dean (Academics), School of Electronics Engineering, Vellore Institute of Technology, Chennai; reenamonica@vit.ac.in

* Correspondence: reenamonica@vit.ac.in

**Abstract:** Cardiac Pacemakers are used to regulate the heart's rhythm and prevent abnormal heart beats. Patients undergo a minimal surgery to get the pacemaker implanted in the body, whereas these devices do have their limitations such as battery life, power consumption and integrated circuit area which makes it difficult for elderly and children to undergo this surgery. This paper focuses on developing an optimised low pass filter ASIC design for implantable QRS complex detector for cardiac pacemaker circuits using filter optimisation techniques such as Pipelining and Folding of filters. The folded low pass filter design consumes an overall power of 0.7575 mW and comprises a total of 1361 standard cells. The number of adder block units in our work reduces area consumption by a factor of 48.37%.

## 1. Introduction

Advancements in medical and technology have led to the development of sophisticated electronic devices that are increasingly being used in the diagnosis and treatment of various medical conditions. One such medical device is the implantable cardiac pacemaker [1] which is a medical device that is implanted in a person's chest to regulate their heart rate and rhythm. It is commonly used to treat arrhythmias, which are irregular heartbeats that can cause dizziness, shortness of breath, and fainting. Pacemakers work by sending electrical signals to the heart to regulate its rhythm and ensure that it beats at a consistent rate. There are several types of pacemakers, including the single-chamber, dual-chamber and bi-ventricular pacemakers. Generally, these pacemaker circuits consists of of a pulse generator that generates the required electrical current for stimulation of heart musculature and one or two electrodes/leads, which are responsible for transmitting these electrical activity to the heart. [2]

The ECG signal waves, also known as electrocardiogram represents the electrical activity of the heart as it contracts and relaxes, and it provides important information about the heart's rhythm and function. It is usually composed of several periodic distinct waves, each of which represents a different phase of the cardiac cycle. The main waves include the P wave, QRS complex, and T wave as represented in Figure 1. The P wave represents the electrical activity that occurs when the atria of the heart contract, while the QRS complex represents the electrical activity that occurs when the ventricles of the heart contract. The T wave represents the electrical activity that occurs when the ventricles of the heart relax and prepare for the next contraction.

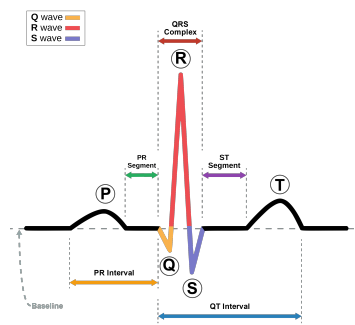

**Figure 1.** ECG Signal Wave.

The frequency range of the P wave and T wave, which are lower frequency components, is 0.5-5Hz. On the other hand, the QRS complex is a higher frequency wave with a range of 0-45Hz. It is evident that the QRS complex has a higher energy value in the R-R interval when compared to the P and T wave. Therefore, to ensure precise monitoring and correction of the heartbeat through electrical signalling, pacemaker circuits must accurately detect the QRS complex without any noise interference. Additionally, to avoid the need for repeated surgery and pacemaker replacements, the battery must be optimised to have a longer lifespan. [3]

Various factors can cause noise interference, with one predominant cause being power line interference, which results from AC signals and has a frequency range of 50-60Hz. Baseline interference affects the lower frequency components, as their frequency range falls between 0.15-0.3Hz and is largely caused by human breathing. Electro-Magnetic field interference occurs due to the electrical nature of ECG and inadequate shielding in the equipment. Poor skin-to-electrode contact can also impact the ECG signal, while other interferences, such as muscle contractions and physical movement, can generate noise with a frequency range of 0-10 kHz.

Suppressing these artefacts is the preliminary step in analysing the ECG Signals and various methods such as Filtering, Hidden Markov Model (HMM) and Wavelet Transformation can be used to achieve this. Filtering the ECG Signal by using low pass, high pass, band pass or band stop filters will result in the removal of lot of wanted signals along with the artefacts as we downscale from 360Hz to 45Hz. HMM, Combinational or probabilistic methods will take the ECG signal very near to ideal conditions but the hardware complexity of the same will be so high resulting in a greater computational time. To avoid this trade off between hardware complexity and denoising, Wavelet transformation is used to suppress the artefacts from the ECG signals. [1]

Hence, the deployment of accurate QRS complex detection without noise interference and low power techniques, without compromising on efficiency and chip size, remains a challenging problem. In this work, we proposed a smaller low power low pass filter design for implantable cardiac pacemaker.

## 2. Review of Literature

The concept of using electrical stimulation to regulate the heartbeat was first proposed in the early 1900s, and since then, cardiac pacemakers have undergone significant evolution. The first external pacemaker was created by Albert Hyman in 1932, while the first implantable pacemaker was proposed by Wilson Greatbatch in 1960. With the advent of transistor technology in the 1960s and 1970s, pacemakers became smaller and more portable. Moreover, the introduction of lithium batteries in the 1970s and 1980s improved the reliability and longevity of pacemakers. In recent years, pacemaker technology has continued to advance, including the development of lead-less pacemakers that don't require wired leads to connect to the heart. [4]

To reduce the computational complexity and power consumption while maintaining accuracy and extending battery life in cardiac pacemakers, an ultra low-energy FFT ASIC is developed using a 90-nm standard cell library and gate level switching. This approach uses the simple concept of hashing and lookup table to effectively reduce the number of arithmetic operations required to perform the FFT of an ECG signal. While a central processor supervises the overall operation, a dedicated co-processor executes the signal processing algorithm, including fast Fourier transform or similar spectrum analysis algorithms [5-7]. Additionally, FFT transform is used to calculate the frequency spectrum of the ECG signal, and decision tree classifier is then applied to detect various types of arrhythmias [8]. To further improve the efficiency of the FFT algorithm, a butterfly structure is employed to parallelise it, and this parallel FFT algorithm is implemented on a field programmable gate array (FPGA) platform, resulting in significantly faster processing times while maintaining accuracy [9].

Traditional pacemaker systems use wired leads to connect the pacemaker to the heart, which can lead to complications like infection, lead fracture, and insulation failure. To overcome these issues, lead-less pacemaker systems have been developed, which require a compact and efficient antenna for wireless communication between the pacemaker and external devices [10]. One approach to achieve this is to use a rectangular patch structure and a parasitic element to create a dual-band resonance, allowing the antenna to operate at both 2.45 GHz and 5.8 GHz [11]. Another option is to design a compact dual-band antenna using a meander line structure and a modified T-shaped feed-line, which allows the antenna to operate at 401 MHz and 433 MHz while reducing its size and improving its efficiency [12]. Another alternative is to develop a wirelessly powered implantable pacemaker with an on-chip antenna, eliminating the need for a battery and reducing the size and weight of the pacemaker. This design utilises a wireless power transfer method to power the pacemaker and an on-chip antenna to receive the power. Results show that the wirelessly powered pacemaker with an on-chip antenna can achieve similar performance to traditional pacemakers while being smaller and lighter [13]. Similarly, a patch antenna can be used to receive power, and a rectifier and voltage regulator can be used to convert the received power to a suitable voltage for the pacemaker [14]. Furthermore, a planar spiral coil antenna can be used to receive power, and a full-wave rectifier and a low-dropout voltage regulator can be used to convert the received power to a suitable voltage for the pacemaker [15].

The latest advancements propose a non-contact method for capturing pacemaker signals. Although existing non-contact methods exist, they tend to be too complex or require close proximity to the pacemaker, leading to interference and signal quality issues. Hence, a new non-contact sensor that employs an RFID antenna to capture the pacemaker signal from a distance can be utilised. [16] To achieve this, an inductive sensing technique captures the magnetic field around the pacemaker, which is then processed by a signal processing algorithm to extract the pacemaker signal.[17] Additionally, an optical sensing technique captures the changes in light intensity produced by the pacemaker signal, which is also processed using a similar signal processing algorithm to extract the pacemaker signal.[18]

### 3. Materials and Methods

The pacemaker circuit block's primary component is the ECG Wavelet Detector, which is responsible for reconstructing the ECG Signal through wavelet transformation techniques. This detector block includes a wavelet decomposer (wavelet filter bank), a QRS Complex detector, and a noise detector block. The wavelet filter bank breaks down the input ECG signal into sub-bands with either mono-phasic or bi-phasic outputs, which are then sent to the QRS Complex Detector. The detector employs a hypothesis test (multi-scaled product) and a soft threshold comparator to estimate the heart beating rate. The noise detector determines the operation mode of the block based on the signal-to-noise ratio of the ECG input signal, thus reducing power consumption. [1]

The multi-scaled product selects the appropriate wavelet filter from the filter bank for ECG Signal reconstruction based on the feedback from the noise detector. The output reinforces the peaks of the QRS Complex while suppressing noise at this stage. The soft threshold comparator uses a variable threshold with a minimum R-R interval of 200ms. When the comparator detects the QRS Complex, it goes high, and the threshold changes to the highest value. This change is then saved with the 200 samples over the T wave of the previous segment. [1]

The Wavelet filter bank consists of a series of wavelet filters with each having a pair of low pass filters and high pass filters for suppressing the artefacts. Since a dyadic process is used for denoising, the low pass and high pass filter designs play a vital role in determining the computational time and delay in the circuit. Figure 2 illustrates the filter design proposed by [1] and the low pass filter can be further optimised to minimise the area and power consumption which invariably reduces the computational time. Optimisation techniques like pipelining and Folding transformation is implemented on filter design and comparison of the results are stated.

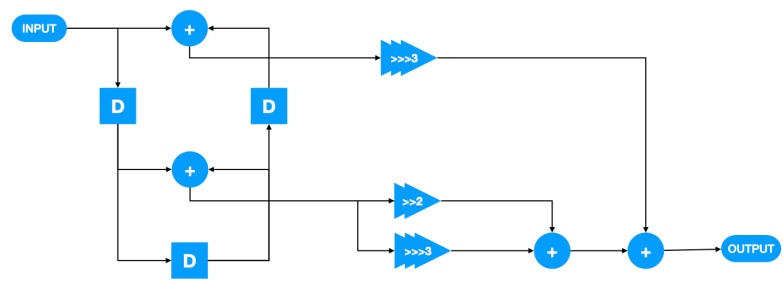

**Figure 2.** Low Pass Filter Design.

Pipelining is a performance optimisation technique that enhances the functioning of digital circuits by fragmenting a complicated task into smaller sub-tasks that can be processed simultaneously. In a pipelined system, the circuit is partitioned into stages, each with a specific function to execute. The output of one stage serves as the input for the next stage, without waiting for the prior stage to finish. This enables several sub-tasks to run concurrently, thereby reducing the overall processing time and accelerating performance. Despite the benefits of pipelining, its application is not without limitations. By introducing more delay units, there is an associated overhead in area and timing analysis, leading to an increase in power consumption and dissipation. Therefore, to mitigate these limitations and optimise circuit performance, Folding followed by register minimisation techniques are implemented to reduce power and area consumption.

The initial step towards implementing pipelining involves identifying the critical path, which is the longest route through the circuit that does not have any delay units. It plays a pivotal role in determining the maximum operating frequency of the circuit. Accurately identifying the critical path is crucial in determining the number of pipeline stages that are necessary to achieve optimal output. Next, the circuit is divided into pipeline stages by identifying the feed forward cut-set. The overall graph is partitioned into stages that are designated to perform specific sub-tasks, and the input and output of each stage are registered to meet the necessary timing requirements. The final step involves adding pipeline registers, which are introduced along the cut-set to store intermediate results and synchronise the timing of signals. Figure 3 depicts the final circuit for the low pass filter with pipeline registers included, with the green delay blocks serving as pipeline registers that split the filter into two pipelined stages along the feed forward cut-set.

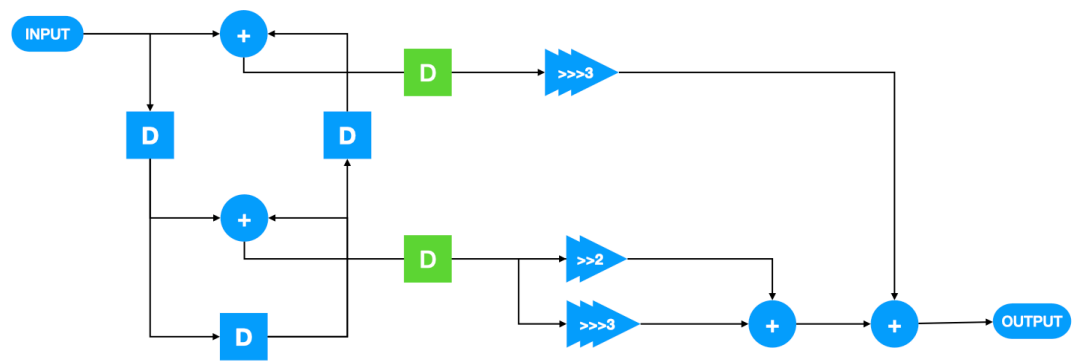

**Figure 3.** Low Pass Filter Pipelined Architecture.

The folding technique is a method that aids in minimising the hardware resources required for implementing digital circuits. This approach involves dividing long interconnection lines into shorter segments that are then folded back onto themselves, allowing for the use of fewer wires and reducing the required chip area for a more efficient and compact design. By reducing wire length and capacitance through the zigzag pattern of folded segments, the folding technique can effectively decrease power consumption and signal delay, making it a widely used technique in modern VLSI design. Additionally, the folding technique enhances the reliability and robustness of integrated circuits, making them more resistant to signal interference and noise. The zigzag pattern reduces the impact of electromagnetic interference that may cause signal distortion and deterioration, leading to improved circuit performance. Ultimately, the folding technique allows more functionality to be packed into a smaller chip area while improving the performance and reliability of the circuit.

To implement folding, the initial step involves identifying the folding factor and folding order. The folding factor refers to the number of segments into which a long interconnect wire is divided and folded back onto itself. This factor determines the degree of zigzagging that occurs in the folded interconnect wire and has a direct impact on routing efficiency, power consumption, and overall circuit performance. Typically, the folding factor is an even number as the folding technique involves symmetrically folding the segments to reduce wire length without introducing any signal delay or glitches in the circuit. On the other hand, the folding order refers to the sequence in which the interconnect wires are folded in a zigzag pattern and determines how the segments are arranged relative to each other. Selecting the appropriate folding factor requires careful consideration of the performance requirements, critical path analysis, resource constraints, and simulation and analysis of the circuit to optimise system performance while minimising area and power consumption. For the low pass filter design in our paper, a folding factor of 2 is employed to fold the design twice, and the folding sets are as follows:

Folding Order (S1) = {A1, A0} and Folding Order (S2) = {A2, A3}.

The next step is evaluating the folding equations, which are used to calculate the delay and power consumption of folded circuits. These equations are derived based on the properties of the folded circuit and the behaviour of individual segments in the folded path. The general form of the folding equation is as follows : $D_F(U - V) = N(w(e)) - P_u + v - u$. Below are the folding equations for the low pass filter circuit discussed in this paper.

$$D_F^a(A_1 - A_2) = 2(1) - 1 + 0 - 0 = 1$$

$$D_F^b(A_1 - A_2) = 2(1) - 1 + 0 - 0 = 1$$

$$D_F(A_2 - A_3) = 2(0) - 1 + 1 - 0 = 0$$

$$D_F(A_0 - A_3) = 2(1) - 1 + 1 - 1 = 1$$

Once the folding equations are applied to determine the delays in each time instance, the final circuit with one adder block is redrawn. If the folding equations result in negative values, timing analysis is required, and the folding sets may need to be reordered to obtain positive values. The final circuit of our design after applying the folding transformation is shown in Figure 4, which includes two yellow adder-delay blocks that represent the folded version of multiple adder units with different time instances.

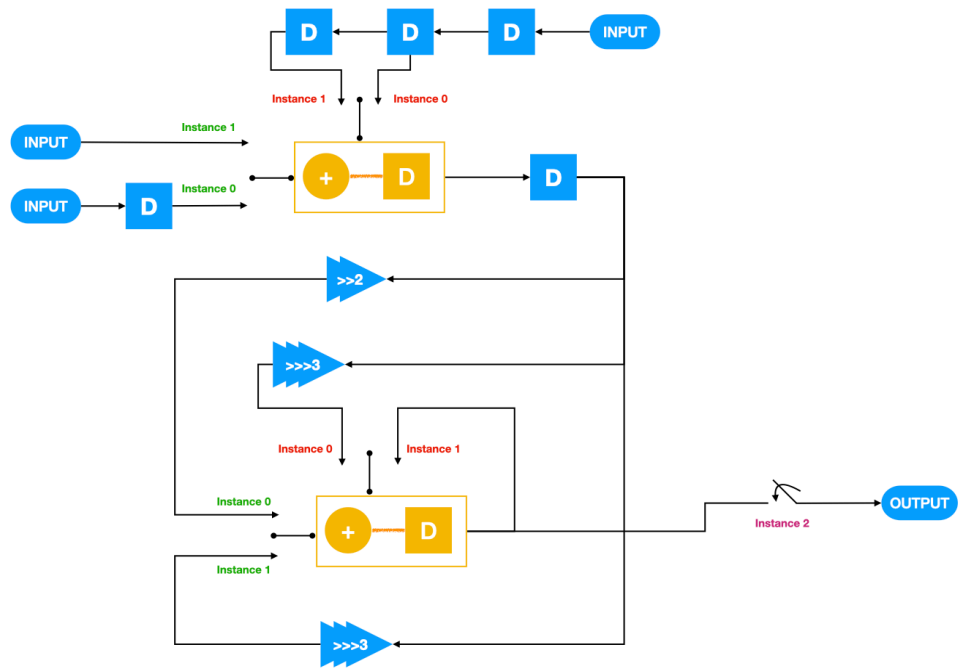

**Figure 4.** Low Pass Filter Folded Architecture

Register minimisation is a technique employed in digital circuit design that aims to maintain the same functionality while reducing the number of registers utilised. Lifetime analysis is the preliminary step in register minimisation which is used to identify the lifespan of registers in a digital circuit. The lifetime of a register is the duration for which the register holds a value before it is overwritten or becomes unused. The process of lifetime analysis involves tracing the flow of data through the circuit and determining when each register is written and read. This information is used to create a lifetime graph, which displays the lifespan of each register in the circuit. Table 1 summarises the lifetime analysis carried out on the filter design and Figure 5 illustrates the lifetime graph.

| Nodes (Adder Block) | $T_{Input}$ to $T_{Output}$ | $T_{Input}$ to $T_{Output}$ |
|---|---|---|
| N1 (A1) | **0+1 to 0+1+0** | 1 - 1 |
| N2 (A0) | **1+1 to 1+1+0** | 2 - 2 |
| N3 (A2) | **0+1 to 0+1+1** | 1 - 2 |
| N4 (A3) | **1+1 to 1+1+1** | 2 - 3 |

**Table 1.** Lifetime Analysis of Low Pass Filter.

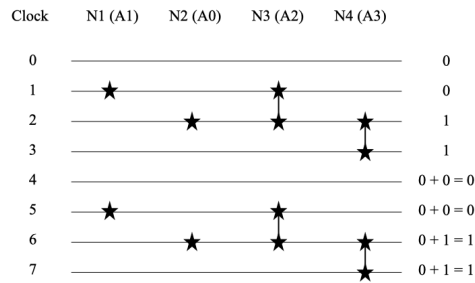

**Figure 5.** Lifetime Graph.

It is evident from the lifetime graph that our low pass filter design needs at least one register to perform the read and write functions separately. However, the folded version of the design has just one register and does not require any additional registers. Therefore, there is no need to minimise the number of registers as the register usage is already optimised.

### 4. Results

The implementation of the design involved three stages. Firstly, the RTL script underwent verification and bit-mapping by running it through the NC Launch Simulation platform, which is a simulation and analysis tool in the Cadence software suite for digital designs. The top module and other modules were compiled, elaborated and simulated, and the input-output correspondence was tested using the top module test bench sent to the waveform.

In the second stage, synthesis was performed using Genus, which is a synthesis tool in the Cadence software suite. The RTL file and constraints file were synthesised to generate a net-list of modules used, and the design's performance was analysed through reports on timing, cell area, power, and delays.

Finally, Innovus, a physical implementation tool in the Cadence software suite, was used to perform the physical design task. The net-list generated by Genus was used to create a layout, and Clock tree synthesis was carried out after placing the cells. The final optimised design was then used to generate reports on power, timing, and area.

Analysing these design reports of the folded low pass filter, it was found that the filter consumes an overall power of 757572.810 nW and comprises a total of 1361 standard cells. Figure 6 gives the ASIC Physical Design layout and Table 2 provides a detailed breakdown of the cell area of the standard cells in the net-list, with a total area of 7493.310 um^2 and an area of 5187.793 um^2 when physical cells are subtracted.

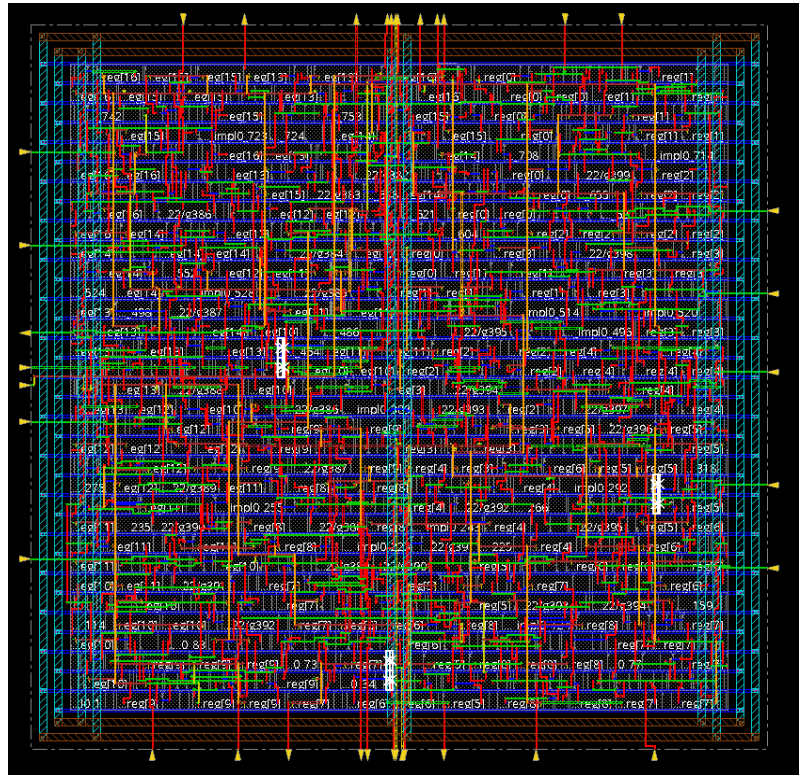

**Figure 6.** ASIC Physical Design layout of the Low Pass Filter.

| Cell Type | Instance Count | Area (um2) |
|---|---|---|
| OAI2BB1X1 | 1 | 5.2983 |
| CLKBUFX2 | 1 | 4.5414 |
| FILL1 | 234 | 177.1146 |
| DFFQX1 | 87 | 1382.8563 |
| AOI2BB1X1 | 1 | 6.0552 |
| AND2X1 | 4 | 18.1656 |
| OAI211X1 | 1 | 5.2983 |
| FILL2 | 226 | 342.1188 |
| XNOR2X1 | 1 | 8.3259 |
| DFFQXL | 107 | 1700.7543 |
| MX2X1 | 62 | 422.3502 |
| FILL8 | 57 | 345.1464 |
| FILL16 | 26 | 314.8704 |
| OAI21X1 | 1 | 4.5414 |
| ADDFX1 | 28 | 551.0232 |
| NAND2XL | 3 | 9.0828 |
| OA21XL | 1 | 6.8121 |
| NOR2BX1 | 227 | 1030.8978 |
| OAI21XL | 1 | 4.5414 |
| FILL4 | 276 | 835.6176 |
| CLKBUFX4 | 4 | 27.2484 |
| FILL32 | 12 | 290.6496 |

**Table 2.** Cell Area of Standard Cell in Net-List.

A comparison between the Cell Area results of the low pass filter design in [1] and the optimised filter design proposed in our work is presented in Table 3.

| Cell Type | Cell Area in [1] (um²) | Cell Area in our work (um²) |
|---|---|---|
| CLKBUF | 6.8121 | 31.7898 |
| FILL | 1279.9179 | 2305.5174 |
| ADD | 1067.229 | 551.0232 |
| AND | 13.6242 | 18.1656 |
| NAND | 0 | 9.0828 |
| NOR | 217.9872 | 1030.8978 |
| XNOR | 0 | 8.3259 |
| DFF | 762.9552 | 3083.6106 |
| MUX | 0 | 422.3502 |
| OAI21 | 4.5414 | 9.0828 |
| AOI2BB1 | 18.1656 | 6.0552 |
| OAI2BB1 | 15.8949 | 5.2983 |
| OAI211 | 10.5966 | 5.2983 |
| SDFFQ | 878.7609 | 0 |
| OA21 | 0 | 6.8121 |

Table 3. Cell Area Comparison.

After examining the data presented, it is clear that despite implementing several optimization techniques in the design, the standard cell count has seen a significant increase of 66.79% from 452 to 1361. However, the cell density has only increased by a factor of 5.27% from 74.362% to 79.633%.

Regarding the number of adder blocks, our work has resulted in a reduction of instance count from 55 to just 28. This decrease has led to a reduction in area consumption from 1,067.229 um^2 to 551.0232 um^2, which represents a 48.37% decrease. Furthermore, the FILL Cell Area has increased by 44.84%, providing more unoccupied space on the silicon area.